\documentclass[fleqn,usenatbib]{mnras}

\usepackage{newtxtext,newtxmath}
\usepackage[T1]{fontenc}
\usepackage{ae,aecompl}

\usepackage{graphicx}	% Including figure files
\usepackage{amsmath}	% Advanced maths commands
\usepackage{amssymb}	% Extra maths symbols

\newcommand{\Ni}{\ensuremath{^{56}\mathrm{Ni}}}

\newcommand{\Msun}{\ensuremath{\mathrm{M}_\odot}}
\newcommand{\Rsun}{\ensuremath{\mathrm{R}_\odot}}
\newcommand{\Msunpyr}{\ensuremath{\mathrm{M}_\odot}~\mathrm{yr^{-1}}}
\newcommand{\vw}{\ensuremath{v_\mathrm{wind}}}
\newcommand{\Mdot}{\ensuremath{\dot{M}}}
\newcommand{\kmps}{\ensuremath{\mathrm{km~s^{-1}}}}

\newcommand{\rhocsm}{\ensuremath{\rho_\mathrm{CSM}}}

\title[Wind acceleration and early SN LCs]{
Immediate dense circumstellar environment of supernova progenitors caused by wind acceleration: its effect on supernova light curves
}

\author[T. J. Moriya et al.]{
Takashi J. Moriya,$^{1}$\thanks{E-mail: takashi.moriya@nao.ac.jp (TJM)}
Sung-Chul Yoon,$^{2}$
G\"otz Gr\"afener,$^{3}$ and
Sergei I. Blinnikov$^{4,5,6}$
\\
$^{1}$Division of Theoretical Astronomy, National Astronomical Observatory of Japan, 
National Institutes of Natural Sciences, \\ 2-21-1 Osawa, Mitaka, Tokyo 181-8588, Japan\\
$^{2}$
Department of Physics and Astronomy, Seoul National University, Gwanak-ro 1, Gwanak-gu, Seoul, 08826, Korea \\
$^{3}$
Argelander Institute for Astronomy, University of Bonn, Auf dem H\"ugel 71, D-53121 Bonn, Germany
\\
$^{4}$
Institute for Theoretical and Experimental Physics, Bolshaya Cheremushkinskaya ulitsa 25, 117218 Moscow, Russia \\
$^{5}$
All-Russia Research Institute of Automatics, Sushchevskaya ulitsa 22, 127055 Moscow, Russia \\
$^{6}$
Kavli Institute for the Physics and Mathematics of the Universe (WPI),
The University of Tokyo Institutes for Advanced Study, \\
The University of Tokyo, 5-1-5 Kashiwanoha, Kashiwa, Chiba 277-8583, Japan \\
}

\date{Accepted 2017 April 13  . Received 2017 April 12 ; in original form 2017 March 08}

\pubyear{2017}

\begin{document}
\label{firstpage}
\pagerange{\pageref{firstpage}--\pageref{lastpage}}
\maketitle

\begin{abstract}
Type~IIP supernova progenitors are often surrounded by dense circumstellar media that may result from mass-loss enhancement of the progenitors shortly before their explosions.
Previous light-curve studies suggest that the mass-loss rates are enhanced
up to $\sim 0.1~\Msunpyr$, assuming a constant wind velocity.  
However, density of circumstellar media at the immediate stellar vicinity can be much higher than previously inferred for a given mass-loss rate if wind acceleration is taken into account.
We show that the wind acceleration has a huge impact when we estimate mass-loss rates from early light curves of Type~IIP supernovae by taking SN~2013fs as an example. We perform numerical calculations of the interaction between supernova ejecta and circumstellar media with a constant mass-loss rate but with a $\beta$-law wind velocity profile. We find that the mass-loss rate of the progenitor of SN~2013fs shortly before the explosion, which was inferred to be $\sim 0.1~\Msunpyr$ with a constant wind velocity of 10~\kmps\ by a previous light-curve modeling, can be as low as $\sim10^{-3}~\Msunpyr$ with the same terminal wind velocity of $10~\kmps$ but with a wind velocity profile with $\beta\simeq 5$. In both cases, the mass of the circumstellar medium is similar $(\simeq 0.5~\Msun)$. Therefore, the beginning of the progenitor's mass-loss enhancement in our interpretation is $\sim 100$ years before the explosion, not several years. Our result indicates that the immediate dense environment of Type~II supernova progenitors may be significantly influenced by wind acceleration.
\end{abstract}

\begin{keywords}
supernovae: general -- supernovae: individual: SN~2013fs -- stars: evolution -- stars: winds, outflows -- stars: mass-loss
\end{keywords}

%%%%%%%%%%%%%%%%%%%%%%%%%%%%%%%%%%%%%%%%%%%%%%%%%%

%%%%%%%%%%%%%%%%% BODY OF PAPER %%%%%%%%%%%%%%%%%%

\section{Introduction}
Type~IIP supernovae (SNe) are the most common type of core-collapse SNe \citep[e.g.,][]{li2011lossii} and they are known to be explosions of red supergiants (RSGs, \citealt{smartt2015review} for a review). Recent large-field and high-cadence transient surveys are starting to catch SNe~IIP within a day after their explosions \citep[e.g.,][]{yaron2017iipcsm,garnavich2016keplerbreakout,quimby2007earlyiip}. These early photometric and spectroscopic properties of SNe~IIP often do not agree with those predicted theoretically. For example, rise times of SNe~IIP are often faster than predicted \citep[e.g.,][]{gonzalez2015iiprise,gall2015earlyiip,rubin2016manyearlyiip} and early SN~IIP spectra also often show narrow lines \citep[e.g.,][]{khazov2016earlyiispectra,smith2015iiniip} as observed in SNe~IIn \citep{schlegel1990iin}. One way to resolve this discrepancy is to assume that there are dense circumstellar media (CSM) surrounding RSG SN progenitors \citep[e.g.,][]{moriya2011iipcsm,nagy2016iipbump,morozova2016iil}.

Recently, SN~2013fs provided a clear case of a RSG SN progenitor surrounded by dense CSM at the immediate vicinity of the progenitor \citep{yaron2017iipcsm}. SN~2013fs was caught within a few hours after the explosion and its first spectrum was taken in about 6 hours after the explosion. It had a flash spectroscopic feature with narrow lines indicating the existence of dense CSM \citep[e.g.,][]{gal-yam2014flash,groh2014flash,grafener2016flashmodel}. However, the narrow lines disappeared in several days and the spectra changed to those of normal SNe~IIP. Light-curve (LC) modeling by \citet{morozova2016iil} also suggested that a dense CSM located at the immediate vicinity of the progenitor is required to explain the early LC properties of SN~2013fs.

The high CSM density in the immediate vicinity of the SN progenitors has been suggested to be caused by an increase of the progenitors' mass-loss rates several years before their explosions.\footnote{Alternatively, the matter ejected by a wind can be confined by a strong external radiation field to make dense CSM \citep{mackey2014pico}.} In the canonical stellar evolution theory, no significant increase in the mass-loss rates in several years before the SN explosions is expected \citep{langer2012review} and several mechanisms to explain the possible mass-loss enhancement has been suggested \citep[e.g.,][]{quataert2012wavemassloss,moriya2014neutrino,heger1997superwind,yoon2010cantiello,moriya2015langer}. A precise estimate of the mass-loss rate shortly before the explosion is essential in understanding the unknown mechanisms of the pre-SN mass loss. In this Letter, we show that the effect of wind acceleration on the CSM density structure has a great impact on early LCs of SNe~IIP. A wind launched at the stellar surface is gradually accelerated to reach the terminal velocity. Therefore, the wind velocity just above the stellar surface is lower than the terminal velocity. This leads to a higher density of the CSM close to the progenitor than in the case where a constant wind velocity is assumed for a given mass-loss rate. Here, we show that SN progenitors' mass-loss rates shortly before their explosions can be significantly overestimated if the effect of wind acceleration is not properly taken into account by taking SN~2013fs as an example.

\section{Progenitor system}\label{sec:progenitor}
\subsection{Progenitor model}
As our focus in this Letter is on SNe~IIP, we adopt a RSG progenitor model obtained with the public stellar evolution code \texttt{MESA} \citep{paxton2011mesa1,paxton2013mesa2,paxton2015mesa3}. The progenitor has a zero-age main-sequence mass of 12~\Msun\ with the solar metallicity. The Ledoux criterion for convection is used with a mixing-length parameter of 2.0 and a semi-convection parameter of 0.01. Overshooting is applied on top of the hydrogen burning convective core with a step function. The adopted overshoot parameter is $0.3~H_P$, where $H_P$ is the pressure scale height at the outer boundary of the convective core. The standard `Dutch' wind scheme is used for both hot and cold winds, with a scaling factor of 1.0. The model has been evolved to the stage of oxygen burning in the core, from which the hydrogen enveolpe structure hardly changes until the core collapse.

The RSG progenitor has 10.3~\Msun\ with the hydrogen-rich envelope of 6.1~\Msun. Its radius is $R_\star=607~\Rsun$ and the effective temperature is 3500~K. The final mass-loss rate of the progenitor according to the `Dutch' wind prescription is $1.4\times 10^{-6}~\Msunpyr$. However, as explained below, we assume that the wind mass-loss rate is enhanced shortly before the SN explosion such that a dense CSM is formed while the progenitor structure is not significantly affected.

\begin{table}
	\centering
	\caption{Wind properties.}
    \label{tab:wind}
	\begin{tabular}{ccccc} % four columns, alignment for each
		\hline
		\Mdot & $v_\infty$  & $v_0$ & $\beta$ & CSM radius \\
        $\Msun~\mathrm{yr^{-1}}$ & \kmps & $\mathrm{m~s^{-1}}$ & & cm \\
		\hline
		$10^{-3}$ & 10 & 10 & 5 & $10^{15}$\\
		$10^{-4}$ & 10 & 1 & 5 & $10^{15}$ \\
		0.15$^a$ & 10 & - & 0 & $1.3\times 10^{14}$ \\
		\hline
        \multicolumn{5}{l}{$^a$ A constant $\vw$ model by \citet{morozova2016iil}.}\\
	\end{tabular}
\end{table}

\begin{figure}
	\includegraphics[width=\columnwidth]{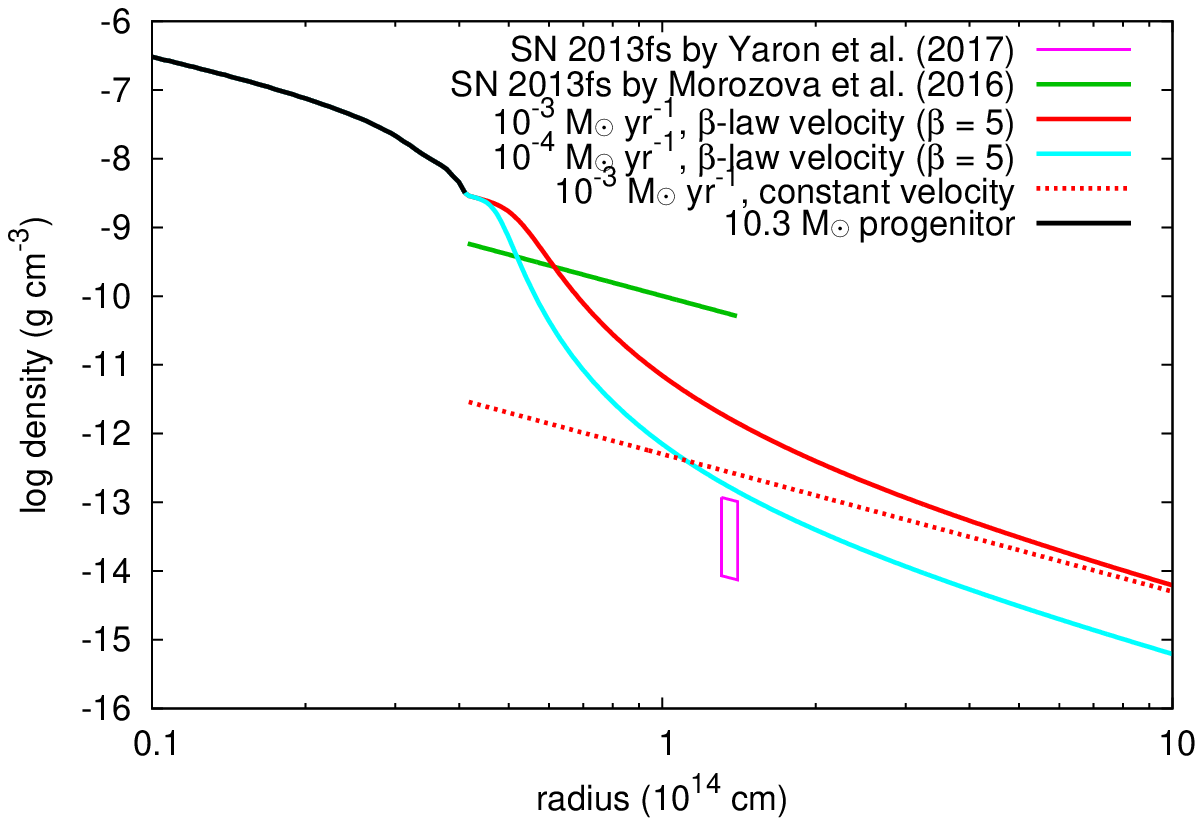}
    \caption{
    Density structure of CSM. Our CSM model has a constant mass-loss rate but the $\beta$-law wind velocity ($\beta=5$). We also show a dashed line where a constant mass-loss rate $(10^{-3}~\Msunpyr)$ and a constant wind velocity $(10~\kmps)$ are assumed.
    The CSM density of SN~2013fs estimated by \citet{yaron2017iipcsm} from spectral modeling is shown with a square.
    The CSM density of SN~2013fs estimated by \citet{morozova2016iil} from early LC modeling is also presented.
    }
    \label{fig:csm}
\end{figure}

\subsection{CSM}
CSM density (\rhocsm) is determined by the progenitor's mass-loss rate (\Mdot) and wind velocity (\vw) as
\begin{equation}
\rhocsm (r) = \frac{\Mdot}{4\pi \vw (r)}r^{-2}.
\end{equation}
In this work, we adopt a $\beta$-law velocity profile as
\begin{equation}
\vw (r) = v_0 + (v_\infty - v_0) \left( 1 - \frac{R_0}{r} \right)^\beta, \label{eq:betavel}
\end{equation}
where $v_0$ is the initial wind velocity at the stellar surface, $v_\infty$ is the final wind velocity, and $R_0$ is where the wind is launched. We find that $\beta \simeq 5$ provides the best fit to the SN~2013fs LC and we focus on this $\beta$ in this Letter.
Observations \citep[e.g.,][]{bennett2010rsgwind,marshall2004agbwindacc} indicate that RSG winds have larger values of $\beta$, i.e., slower wind acceleration, than OB supergiants that have $\beta\simeq 0.5-1$ \citep[e.g.,][]{groenewegen1989ostarmassloss,haser1995ostarmassloss,puls1996ostarmassloss}. For example, the $\beta$-law velocity profile with $\beta\simeq 3.5$ is found to match a RSG $\zeta$ Aurigae \citep{baade1996wind}. We also set $R_0=R_\star$ in this work assuming that the wind is launched at the stellar surface. $R_0$ can vary depending on the wind acceleration mechanisms \citep[e.g.,][]{bennett2010rsgwind,lamers1999windbook}. For example, $R_0$ can be $2-3~R_\star$ in the dust-driven winds. We found that models with $R_0 = R_\star$ match SN~2013fs and we focus on this case in this Letter. We investigate the effects of different $\beta$ and $R_0$ in our future work.

We present LCs for two different mass-loss rates: $10^{-3}$ and $10^{-4}~\Msunpyr$ (Table~\ref{tab:wind} and Fig.~\ref{fig:csm}). We set the terminal wind velocity $v_\infty=10~\kmps$. The initial velocity $v_0$ is chosen so that the density structure at the stellar surface is smoothly connected. We cut our dense CSM at an arbitrary radius of $10^{15}~\mathrm{cm}$. 

Fig.~\ref{fig:csm} shows the density structure of our models. We also present the density structure that gives the best fit to the early LCs of SN~2013fs assuming a constant \Mdot\ and \vw\ \citep{morozova2016iil}. Our CSM have similar density to those found by \citet{morozova2016iil} near the progenitor, but our mass-loss rates are only $10^{-3}-10^{-4}~\Msunpyr$ while $0.15~\Msunpyr$ is required when the constant wind velocity of 10~\kmps\ is assumed. The CSM mass in our $10^{-3}~\Msunpyr$ model is $0.5~\Msun$ while that in the model of \citet{morozova2016iil} is $0.4~\Msun$.

\begin{figure*}
	\includegraphics[width=\columnwidth]{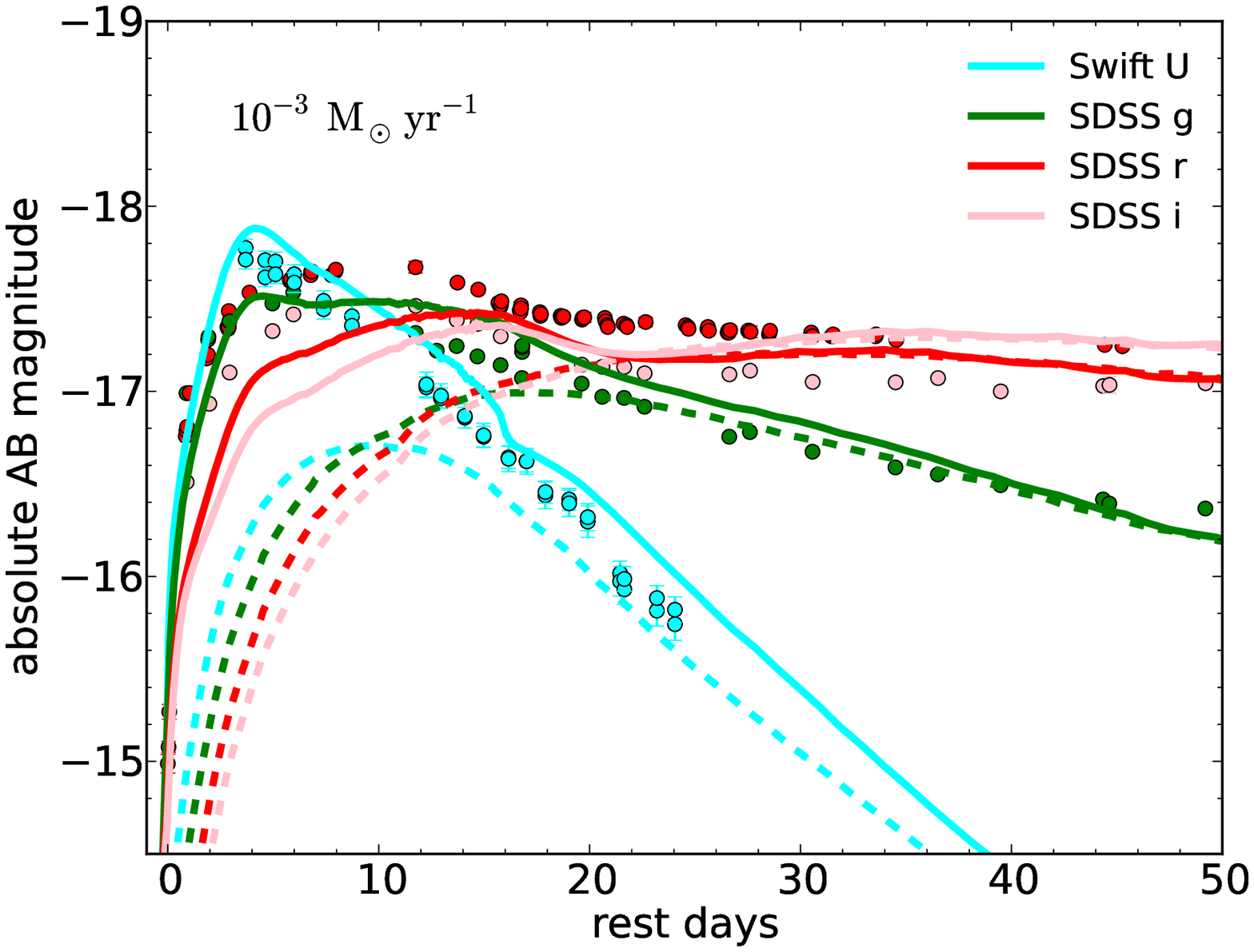}
    \includegraphics[width=\columnwidth]{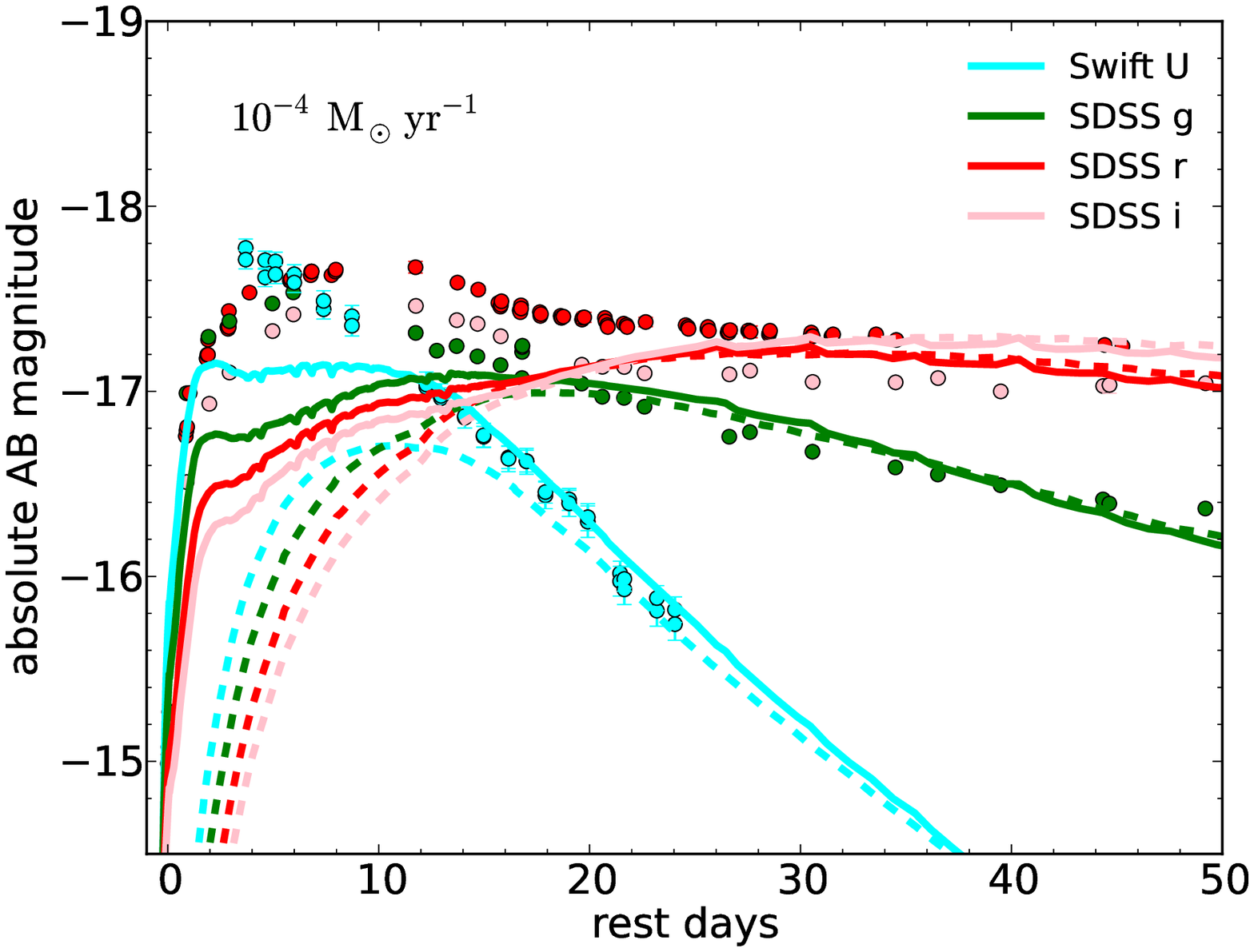}
    \caption{
    Synthetic LCs from our RSG progenitor with an explosion energy of $10^{51}~\mathrm{erg}$ and LCs of SN~2013fs. The left panel shows our model with $10^{-3}~\Msunpyr$ and the right shows that with $10^{-4}~\Msunpyr$. The LC models with dashed lines are those without CSM.
	}
    \label{fig:lcs_wcsm}
\end{figure*}

\begin{figure}
    \includegraphics[width=\columnwidth]{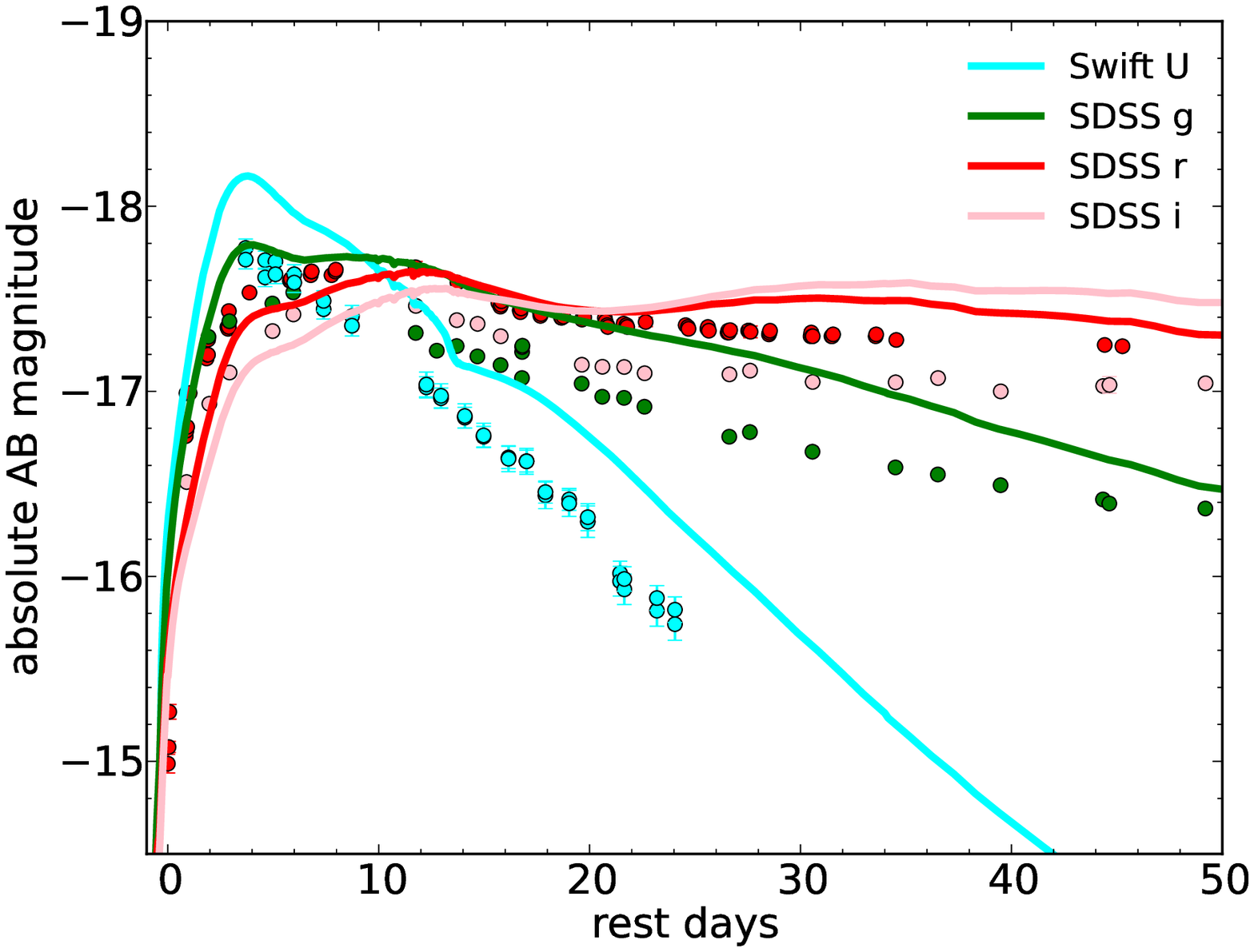}
    \caption{Synthetic LCs from our $10^{-3}~\Msunpyr$ model with the explosion energy of $1.3\times 10^{51}~\mathrm{erg}$.}
    \label{fig:lcs_highe}
\end{figure}

\begin{figure}
    \includegraphics[width=\columnwidth]{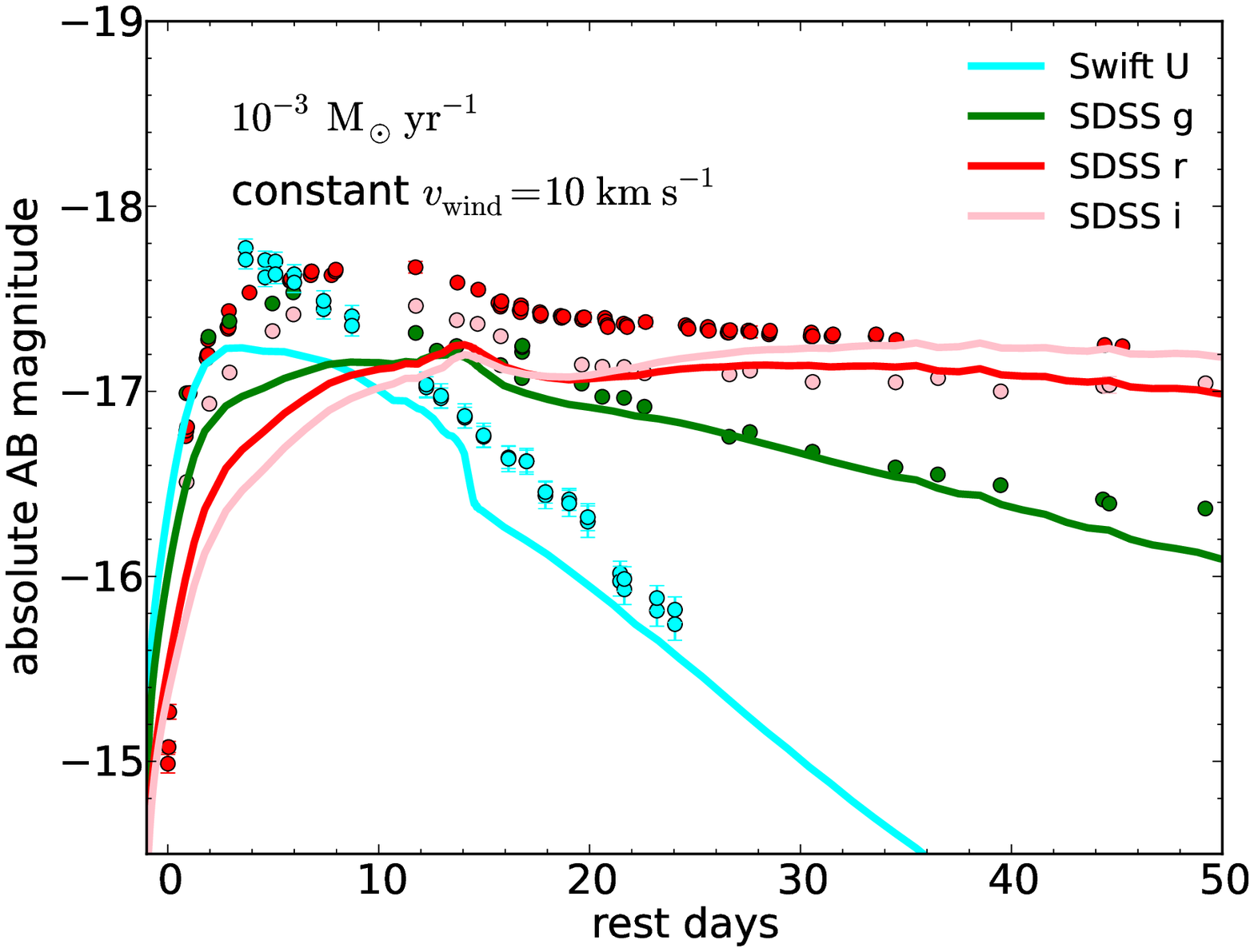}
    \caption{Synthetic LCs from a $10^{-3}~\Msunpyr$ model with a constant wind velocity of $10~\kmps$ without the wind acceleration. An explosion energy of $10^{51}~\mathrm{erg}$ is adopted.}
    \label{fig:1e-3const}
\end{figure}

\section{Light curves}\label{sec:lcs}
\subsection{Methods}
We use a one-dimensional multi-group radiation hydrodynamics code \texttt{STELLA} to investigate LCs numerically \citep[e.g.,][]{blinnikov1998sn1993j,blinnikov2000sn1987a,blinnikov2006sniadeflg,chugai2002windacc}. Starting from the progenitor models presented in the previous section, we initiate explosions by putting thermal energy just below the mass cut that we set at $1.4~\Msun$. We put $0.1~\Msun$ of \Ni\ at the center, but it does not affect the early LCs we present here.

\texttt{STELLA} evaluates spectral energy distributions (SEDs) at each time step. We obtain multicolor LCs by convolving filter transmission functions to the SEDs. Here, we adopt the $U$ band filter of Swift/UVOT \citep{poole2008swiftuvot} and the $g$, $r$, and $i$ filters of the Sloan Digital Sky Survey (SDSS, \citealt{doi2010sdss}). To compare with SN~2013fs, we apply a redshift correction with $z=0.0119$ and a Galactic extinction of $E(B-V)=0.0346$ assuming the extinction curve of \citet{cardelli1989exti} with $R_V =3.1$ to our theoretical SEDs.

The multicolor LCs of SN~2013fs, with which we compare our models, are presented by \citet{valenti2016typeii} and \citet{yaron2017iipcsm}. We acquire machine-readable data through the Open Supernova Catalog \citep{guillochon2017osc}. We take the $U$, $g$, $R$, and $i$-band data for comparison to cover a wide spectral range. Although we use the SDSS $r$-band filter to obtain multicolor LCs from our theoretical SEDs and compare them with the $R$-band observations, the differences between the observed $R$-band and the $r$-band magnitudes are negligibly small \citep{yaron2017iipcsm}.

\subsection{SN~2013fs}
Fig.~\ref{fig:lcs_wcsm} shows LCs obtained from our CSM configuration. We adopt the explosion energy of $10^{51}~\mathrm{erg}$ to match the luminosity during the later plateau phase. The dashed lines in the figure are LC models without CSM. The existence of the dense CSM can explain the early LCs of SN~2013fs as presented by \citet{morozova2016iil}.

We find that our LC model with the CSM having $\Mdot=10^{-3}~\Msunpyr$ provides a LC that matches SN~2013fs reasonably well. The LC model with $\Mdot=10^{-4}~\Msunpyr$ also shows a fast rise but it does not become as bright as SN~2013fs. Although the $U$ and $g$-band LCs of SN~2013fs are well reproduced by the $\Mdot=10^{-3}~\Msunpyr$ model, it is slightly fainter than SN~2013fs in the $r$ and $i$ bands by $\sim 0.4~\mathrm{mag}$. The $r$ and $i$-band brightness can be matched to SN~2013fs by increasing the explosion energy to $1.3\times 10^{51}~\mathrm{erg}$, but then the $U$ and $g$-band magnitudes becomes brighter by $\sim 0.4~\mathrm{mag}$. This slight difference may be caused by, e.g., incompleteness of opacity information we adopt or higher metallicity of the progenitor of SN~2013fs. LCs from a constant mass-loss rate of $10^{-3}~\Msunpyr$ and a constant wind velocity of 10~\kmps\ without the wind acceleration do not become bright enough to explain the early LCs (Fig.~\ref{fig:1e-3const}).

A fundamental difference between the model of \citet{morozova2016iil} and ours is in the mass-loss rates. \citet{morozova2016iil} find that an extremely high mass-loss rate ($\gtrsim 0.1~\Msunpyr$) is required to explain the early LC of SN~2013fs by assuming a constant wind velocity of 10~\kmps. However, we find that a CSM from $10^{-3}~\Msunpyr$ with the terminal velocity of 10~\kmps\ can provide a reasonable fit to the early LC if we take the velocity change due to the wind acceleration into account. Although the mass-loss rate and wind velocity are different in the two models, the final CSM mass is comparable in them ($0.5~\Msun$ in our $10^{-3}~\Msunpyr$ model and $0.4~\Msun$ in \citealt{morozova2016iil}). Therefore, the CSM mass is likely better constrained by the early LCs.

\citet{yaron2017iipcsm} estimate the CSM density at the immediate vicinity of the progenitor of SN~2013fs by modeling their flash spectra. We show their constraint on the CSM density with a square in Fig.~\ref{fig:csm}. Our model with $10^{-4}~\Msunpyr$ is consistent with their estimates but it does not result in a LC that is bright enough to explain the early LC. Our best LC model with $10^{-3}~\Msunpyr$ has higher density by a factor of 10 and the model obtained by \citet{morozova2016iil} has even higher density. \citet{yaron2017iipcsm} may have underestimated $\dot{M}$ by several factors because of the neglect of light-travel-time effects in their study \citep{grafener2016flashmodel}. Together with the altered density structure, this may bring our results in better agreement with their density estimates.

\section{Discussion and conclusions}
\citet{yaron2017iipcsm} interpret that the high CSM density needed to explain the early SN properties of SN~2013fs is caused by the increase of the progenitor's mass-loss rate shortly before the explosion. The early phase LC modeling of \citet{morozova2016iil} indicated that the progenitor's mass-loss rate may have been as high as $0.15~\Msunpyr$ assuming a constant wind velocity of 10~\kmps. However, we have shown that only $\sim 10^{-3}~\Msunpyr$ is sufficient to explain the early LC if we consider wind acceleration. The inferred mass-loss rate of $\sim 10^{-3}~\Msunpyr$ is still very high compared to those of observed RSGs \citep[e.g.,][]{mauron2011rsgmassloss,goldman2017agbrsgwind}. This high mass-loss rate of $\sim 10^{-3}~\Msunpyr$ could not have been maintained for more than $\sim 10^3$~years before the SN explosion, as otherwise most of the hydrogen envelope would have been stripped off. The plateau duration of SN~2013fs ($\simeq 80~\mathrm{days}$) implies that its progenitor had a massive hydrogen-rich envelope. 
In addition, it takes 500~years to reach $10^{14}~\mathrm{cm}$ and then only 50 years to reach $10^{15}~\mathrm{cm}$ in our wind acceleration model with $10^{-3}~\Msunpyr$. \citet{yaron2017iipcsm} found that the dense CSM only extends up to $\simeq 10^{15}~\mathrm{cm}$ in SN~2013fs. Therefore, the mass-loss enhancement should not last more than about 550 years in order not to have too extended dense CSM. On the contrary, if the mass-loss enhancement lasts less than $\sim 100~\mathrm{years}$, the dense CSM does not extend enough to affect the early SN properties. Therefore, the progenitor must have undergone an abrupt increase of mass loss starting at around 500~years before the explosion in our model.

As the estimated CSM mass in SN~2013fs is similar ($\simeq 0.5~\Msun$) in the constant wind velocity model and the wind acceleration model, the difference in the estimated mass-loss rates makes a significant difference in the estimates for the period of the mass-loss enhancement before the explosion. In our model with $\sim 10^{-3}~\Msunpyr$, the mass-loss enhancement is estimated to occur in the final 500~years to the explosion. 
If we assume $\sim 0.1~\Msunpyr$, the mass-loss enhancement should only occur in the final several years to the explosion. These estimates are important in constraining possible mass-loss mechanisms like wave-driven mass loss \citep[e.g.,][]{shiode2014wavemassloss}.

It is possible that the dense CSM formed by the mass-loss enhancement affects not only the early SN properties but also observational properties of underlying RSG SN progenitors. The effects on the observational properties of the progenitors strongly depend on the opacity in the dense CSM. If the CSM has a temperature similar to the RSG photosphere, the opacity is expected to be $\sim 10^{-3}~\mathrm{cm^2~g^{-1}}$ at most \citep{ferguson2005lowtempopacity} and it does not affect the RSG properties significantly. However, the unknown mechanisms of the mass-loss enhancement can strongly affect the opacity in the CSM. For example, if the opacity in the dense CSM with $10^{-3}~\Msunpyr$ becomes $0.01~\mathrm{cm^2~g^{-1}}$, the photosphere locates at $4~R_\star$ and the effective temperature is altered to about 1800~K.
Moreover, if the mass-loss enhancement leads to a formation of a large amount of dusts, the progenitor would be enshrouded by dusts and significantly reddened. Then, if the mass-loss enhancement starts $\sim 100$~years before the explosion in some RSG SN progenitors as we suggest, some of detected RSG SN progenitors so far may have been significantly affected by absorption in the CSM and their progenitor masses may have been estimated lower than they are, for example \citep[e.g.,][]{walmswell2012dustrsgprobsol,beasor2016dustyrsg}.

The observational data of early LCs of SNe~IIP have dramatically increased during the last several years. They give evidence that not a small fraction of SNe~IIP progenitors may be surrounded by immediate dense CSM \citep[e.g.,][]{gonzalez2015iiprise}, like in the case of SN~2013fs. The present study indicates that the effect of wind acceleration should not be ignored in the analysis of early SN~IIP LCs. Addressing this issue with more details including spectral properties is required to better understand the mass-loss history of their progenitors. We also note that the wind acceleration itself is not known well. For example, the shape of the early SN LCs may be affected by the CSM structure and early SN observations may also be useful to constrain wind acceleration such as $\beta$.

\section*{Acknowledgements}
We thank the referee, James Fuller, for constructive comments that improved this work.
TJM thanks the Yukawa Institute for Theoretical Physics at Kyoto University, where this work was initiated during the YITP-T-16-05 on "Transient Universe in the Big Survey Era: Understanding the Nature of Astrophysical Explosive Phenomena". TJM is supported by the Grant-in-Aid for Research Activity Start-up of the Japan Society for the Promotion of Science (16H07413). SCY is supported by the Korea Astronomy and Space Science Institute under the R\&D program (Project No. 3348- 20160002) supervised by the Ministry of Science, ICT and Future Planning. GG is supported by the Deutsche Forschungsgemeinschaft, Grant No. GR 1717/5. Grant no. IZ73Z0 152485 SCOPES Swiss National Science Foundation supports work of SIB. Numerical computations were partially carried out on PC cluster at Center for Computational Astrophysics, National Astronomical Observatory of Japan.

%%%%%%%%%%%%%%%%%%%%%%%%%%%%%%%%%%%%%%%%%%%%%%%%%%



\bibliographystyle{mnras}
\bibliography{ms} % if your bibtex file is called example.bib

%\begin{thebibliography}{99}
%\end{thebibliography}

%%%%%%%%%%%%%%%%%%%%%%%%%%%%%%%%%%%%%%%%%%%%%%%%%%

%%%%%%%%%%%%%%%%% APPENDICES %%%%%%%%%%%%%%%%%%%%%

%\appendix


%%%%%%%%%%%%%%%%%%%%%%%%%%%%%%%%%%%%%%%%%%%%%%%%%%

% Don't change these lines
\bsp	% typesetting comment
\label{lastpage}
\end{document}